\documentclass[
aps,%
12pt,%
final,%
notitlepage,%
oneside,%
onecolumn,%
nobibnotes,%
nofootinbib,%
superscriptaddress,%
noshowpacs,%
centertags]%
{revtex4}

\newcommand{\beq}{\begin{equation}}
\newcommand{\eeq}{\end{equation}}
\newcommand{\bea}{\begin{eqnarray}}
\newcommand{\eea}{\end{eqnarray}}
\newcommand{\Dsl}{D\hspace{-2.4mm}/}

\usepackage{graphicx}

\begin{document}
\selectlanguage{english}

\title{Effects of mesonic correlations in the QCD phase transition}

\author{D. Blaschke}
\email{blaschke@ift.uni.wroc.pl}
\affiliation{Institute for Theoretical Physics, University of Wroclaw,
50-204 Wroclaw, Poland}
\affiliation{Bogoliubov  Laboratory of Theoretical Physics, JINR Dubna,
141980  Dubna, Russia}

\author{M. Buballa}
\email{michael.buballa@physik.tu-darmstadt.de}
\affiliation{Institut f\"ur Kernphysik, Technische Universit\"at Darmstadt,
D-64289 Darmstadt,
Germany}

\author{A.~E.~Radzhabov}
\email{aradzh@theor.jinr.ru}
\affiliation{Bogoliubov  Laboratory of Theoretical Physics, JINR Dubna,
141980  Dubna, Russia}

\author{M.~K.~Volkov}
\email{volkov@theor.jinr.ru}
\affiliation{Bogoliubov  Laboratory of Theoretical Physics, JINR Dubna,
141980  Dubna, Russia}

\begin{abstract}
The finite temperature phase transition of strongly interacting matter is
studied within a nonlocal chiral quark model of the NJL type coupled to a
Polyakov loop.
In contrast to previous investigations which were restricted to the mean-field
approximation, mesonic correlations are included by evaluating the
quark-antiquark ring sum.
For physical pion masses, we find that the pions dominate the pressure below
the phase transition, whereas above $T_c$ the pressure is well described by
the mean-field approximation result.
For large pion masses, as realized in lattice simulations, the meson
effects are suppressed.
\end{abstract}
\maketitle
The interest in the phase diagram of quantum chromodynamics (QCD)
has received new impulses from the recent results of heavy-ion
collision experiments at RHIC Brookhaven \cite{Muller:2006ee},
revealing that hot hadronic matter
at temperatures not too far above the critical temperature $T_c$
behaves like a perfect liquid, rather than a weakly coupled plasma.
A theoretical description of these investigations therefore requires
a non-perturbative approach, which also provides a proper understanding
of the chiral quark dynamics and the confinement mechanism.
Until now, the only method which is directly based on QCD and which
meets these requirements is lattice gauge theory.
Unfortunately, the application of lattice results to experimental data is
complicated by the fact that most lattice calculations are
performed with rather large quark masses, leading to unphysically
large pion masses.
Therefore, in spite of recent progress in this aspect,
additional methods are required to extrapolate the lattice results
to the physical masses.
To bridge the gap there are major efforts within chiral
perturbation theory ($\chi PT$) (see, e.g., \cite{Meissner:2006ai}).
However, $\chi PT$ is obviously not suited to describe the region of the phase
boundary or even beyond.
In this regime, one has to employ other approaches for the chiral
extrapolation, like QCD Dyson-Schwinger equations
(see, e.g., \cite{DSE}) or effective models which share
the relevant degrees of freedom with nonperturbative QCD
\cite{Bender:1996bm}.
These approaches have the additional advantage that they can straightforwardly
be extended to nonzero chemical potential, which is still
a serious problem on the lattice.

In this article we want to discuss an effective model of low-energy QCD,
capable of describing the chiral as well as the deconfinement transitions.
As a basis we use the PNJL model
\cite{Ratti:2005jh,PNJL,Hansen:2006ee,Roessner:2006xn} which generalizes
the well-known Nambu--Jona-Lasinio (NJL) model \cite{NJL} for the chiral quark
dynamics by coupling it to the Polyakov loop, which serves as an order
parameter of the deconfinement transition.
To a large extent, this removes one of
the most disturbing features of the original NJL model, namely the pressure
contribution of unconfined quarks in the hadronic phase.
In spite of the simplicity of the model a remarkable agreement
with the results of lattice QCD thermodynamics \cite{Allton:2003vx,AliKhan:2001ek} has
been obtained \cite{Ratti:2005jh}.

However, this comparison was not entirely consistent:
Whereas unphysically large values of the
current quark masses have been
used in the lattice simulations \cite{Allton:2003vx,AliKhan:2001ek},
physical values have been employed in the
PNJL model analysis of Ref.~\cite{Ratti:2005jh}.
Moreover, after successfully removing (most of) the unphysical quark degrees
of freedom from the confined phase, the PNJL model treated in mean-field
approximation as in Ref.~\cite{Ratti:2005jh} does not contain {\it any}
degree of freedom in this regime.
Obviously, this is a rather poor description of the hadronic phase at finite
temperature where mesons are expected to become relevant.
Hence, the good agreement of the PNJL results with the lattice data could
partially be accidental in that mesonic correlations have been neglected in
the PNJL analysis, while in the lattice calculations they are suppressed by
the large current quark masses.
To get a consistent picture it is thus important to go beyond the mean-field
approximation and to include mesonic correlations.
This can be done systematically within the framework of a $1/N_c$ expansion
scheme \cite{'tHooft:1973jz}, which has successfully been applied to the NJL
model, e.g., in Refs.~\cite{NJLNc}.

In the present work we suggest a $1/N_c$ improvement of the PNJL model
which is necessary to
disentangle hadronic contributions in the vicinity of and below the
chiral/deconfinement phase transition.
Moreover, the existence of bound states above $T_c$ may be crucial
for the understanding of the properties of the strongly coupled
quark-gluon plasma \cite{Shuryak:2004tx}.
A consistent inclusion of hadron gas contributions should thus include the
dissociation of hadrons being bound states of quarks and antiquarks below the
transition into resonant continuum correlations somewhere above it.
This hadronic Mott-transition \cite{Mott:1968} has been discussed first in
Refs. \cite{hMott} and was formulated within the NJL
model in Ref. \cite{NJLMott} employing the concept of spectral functions and
in-medium scattering phase shifts within a Beth-Uhlenbeck approach,
see \cite{schmidt:1990}.

In this letter, we restrict ourselves to the case of zero chemical potential.
The central quantity for our analysis is then the thermodynamic potential at
finite temperature which can be written as a sum of a mean-field part and a
part which describes the mesonic correlations,
\beq
\label{omega}
    \Omega(T) = \Omega_\mathrm{mf}(T) + \Omega_\mathrm{corr}(T)- \Omega_0~.
\eeq
As usual, we have introduced an irrelevant constant $\Omega_0$, which is chosen
such that $\Omega(0)=0$.
Technically, the correlation part $\Omega_\mathrm{corr}(T)$ corresponds to
a ring-sum of quark-antiquark loops (see \cite{NJLMott,Ripka:1997zb}).
For the evaluation of (\ref{omega}), we consider a nonlocal generalization of
the PNJL model.
The nonlocal four-point interaction is chosen in a separable form motivated
by the instanton liquid model (ILM) \cite{ILM} where it results
from the internal nonlocal structure of the nonperturbative QCD vacuum.
In the ILM, the nonlocality is represented by the profile function of the
quark zero-mode in the instanton field and depends on the gauge.
We use a  Gaussian ansatz as one of the simplest functional forms of the
nonlocality which has a similar behavior as the zero-mode profile obtained in
a gauge invariant manner \cite{Dorokhov:2000gu}.
This choice guarantees convergence at all orders without an additional
regularization procedure.

The quark sector of the nonlocal chiral quark model is described by the
Lagrangian
\begin{eqnarray}
\mathcal{L}_q&=& \bar{q}(x)(i \Dsl -m_c)q(x)
+\frac{G}{2}[J_\sigma^2(x) + \vec J^{\,2}_\pi(x)] ~,
\end{eqnarray}
where $m_c$ is the current quark mass, and $D_\mu=\partial_\mu-iA_\mu$
the covariant derivative
with a background gluon field
$A_\mu \equiv A^a_\mu\frac{\lambda^a}{2} = \delta_{\mu 0} A_0$.
The nonlocal quark currents are
\begin{eqnarray}
J_I(x)\!=\!\!\int\!\!d^4x_1
d^4x_2\,f(x_1)f(x_2)\,\bar{q}(x-x_1)\,\Gamma_I\,q(x+x_2),
\end{eqnarray}
with the vertices $\Gamma_\sigma=\mathbf{1}$ and
$\Gamma_\pi^a=i \gamma^5 \tau^a$, $a=1,2,3$.

After linearization of the four-fermion vertices by introducing auxiliary
scalar  ($\tilde{\sigma}$) and pseudoscalar ($\pi^a$) meson fields the quark
sector is described by the Lagrangian
\begin{eqnarray}
   \mathcal{L}_{q\pi\sigma} &=& \bar{q}(x)(i \Dsl -m_c)q(x)
   - \frac{\pi_a^2+\tilde{\sigma}^2}{2G}
   + J_\sigma(x)\tilde{\sigma}(x)+\pi^a(x)J_\pi^a(x)~.
\label{lagr2}
\end{eqnarray}
To proceed, we single out the nonzero mean-field value of the scalar field by
the decomposition $\tilde{\sigma}=\sigma+\sigma_{\rm mf}$ so that $\pi^a$ and
$\sigma$ denote only the fluctuating parts of the fields
($\langle\pi^a \rangle=\langle \sigma\rangle=0$)
describing mesonic correlations. The scalar mean field gives a dynamical
contribution to the quark mass which in Euclidean momentum space is
$M(p^2)=m_c+ m_d f^2(p^2)$ with $f^2(p^2)=\exp(-p^2/\Lambda^2)$ being the
(Fourier transformed) Gaussian form factor in Euclidean space.
The amplitude $m_d=-\sigma_{\rm mf}$ is an order parameter for dynamical chiral
symmetry breaking.
The model parameters of the quark sector are fixed by using the pion mass and
the weak pion decay constant and choosing a chiral condensate value in
accordance with the limits from QCD sum rule analyses.
The chiral condensate is obtained from the non-perturbative part of the quark
propagator,
$S_{np}(p)=(p\hspace{-1.8mm}/ + M(p^2))^{-1} - (p\hspace{-1.8mm}/ + m_c)^{-1}$,
i.e., after subtracting the perturbative part.
For details see, e.g., \cite{GomezDumm:2006vz,Birse}.
For the analyses in the present paper, we employ a parameter set from Ref.
\cite{GomezDumm:2006vz} for $-\langle\bar q q \rangle^{1/3}=240$ MeV
given by $m_c=5.8$ MeV, $\Lambda=902.4$ MeV and $G\Lambda^2=15.82$.

The mean-field thermodynamic potential reads
\begin{eqnarray}
\Omega_\mathrm{mf}(T) &=& - 4 \sum\limits_{i=0,\pm} \;\int\limits_{k,n}
\log \left[(\omega_n^i)^2 +\vec{k}^2 + M^2((\omega_n^i)^2) \right]
+ \frac{m_d^2}{2 G} + {\cal U}(\Phi,\bar{\Phi})  ~,
\end{eqnarray}
where the notation $\int_{k,n}=T \sum_n \int {d^3 k}/{(2\pi)^3}$ has been
introduced.
$\Phi$ denotes the Polyakov loop variable, which
is given by $\Phi = \frac{1}{3}\,\mathrm{Tr}_c\,e^{i\phi/T}$.
Here $\phi \equiv A_4 = iA_0$ is related to
the (Euclidean) background gauge field.
In Polyakov gauge it is diagonal in color space, i.e.,
$\phi = \phi_3\lambda_3 + \phi_8\lambda_8$.
Following Ref.~\cite{Roessner:2006xn}, we require
$\Phi = \bar\Phi$ to be real with real $\phi_3,\phi_8$. As a consequence
$\phi_8=0$ and we are left with one variable $\phi_3$
\cite{Roessner:2006xn}.

Due to the coupling to the Polyakov loop the fermionic Matsubara frequencies
$\omega_n=(2n+1)\pi T$ are shifted:
\begin{eqnarray}
\omega_n^\pm=\omega_n \pm \phi_3 , \quad \omega_n^0=\omega_n.
\end{eqnarray}
For the  Polyakov loop potential ${\cal U}(\Phi,\bar{\Phi})$ we adopt the the
logarithmic form of Ref.~\cite{Roessner:2006xn}, which has been
fitted to the quenched lattice data of Ref.~\cite{Kaczmarek:2002mc}.

The order parameters (mean field values for $m_d$ and $\phi_3$) are obtained
by minimization of the mean-field part of the thermodynamic potential
$\frac{\partial \Omega_\mathrm{mf}}{\partial m_d}=0$,
$\frac{\partial \Omega_\mathrm{mf}}{\partial \phi_3}=0$.

\begin{figure}[h]
\setcaptionmargin{5mm}
\onelinecaptionsfalse 
\resizebox{.4\textwidth}{!}{\includegraphics{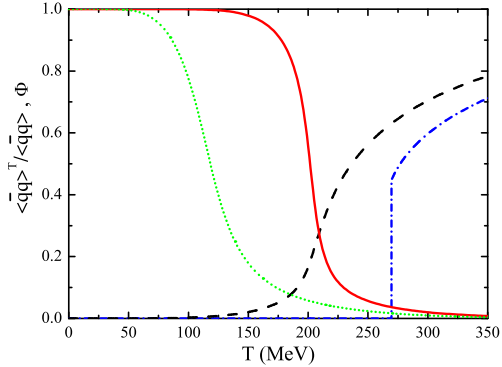}}
\captionstyle{normal} \caption{
Quark condensate (normalized to its vacuum value) in the nonlocal NJL
model (green dotted line) and the nonlocal PNJL model (red solid line) compared
to the Polyakov loop in pure gauge theory (blue dash-dotted line) and in the
nonlocal PNJL model (black dashed line).}
\label{DynMF}
\end{figure}

In Fig. \ref{DynMF} we show the resulting temperature dependence of the quark
condensate $\langle\bar{q}q\rangle^T$ (normalized to its
vacuum value) and of the Polyakov loop expectation value.
We also show the corresponding results for the case of
uncoupled (pure) quark and gluon sectors.
In the pure quark model the critical temperature for chiral restoration is
$T_c=116$ MeV, whereas the pure gauge sector has a critical temperature for
deconfinement $T_d=270$ MeV, fixed from lattice data for the Polyakov loop.
When the quark and gluon sectors are coupled these temperatures get
synchronized so that $T_c\approx T_d \approx 200$ MeV.

We now include $\Omega_\mathit{corr}$ to
study the effect of mesonic correlations.
The properties of mesons in the PNJL model have been studied in
Ref.~\cite{Hansen:2006ee} and the generalization to our nonlocal model
is straightforward. Key ingredients are the quark-antiquark
polarization loops $\Pi_M$.
Using the notations $D(k^2)=k^2+M^2(k^2)$,
${k}_{n+}^i=(\omega_n^i+\nu_m,\vec{k}+\vec{p})$,
and ${k}_{n}^i=(\omega_n^i,\vec{k})$ they can be written as
\begin{eqnarray}
\Pi_{\pi,\sigma}(\vec{p},\nu_m) &=&4 N_f  \sum\limits_{i=0,\pm} \int_{k,n}
\frac{f^2((k_{n+}^i)^2) f^2((k_n^i)^2)}{ D((k_{n+}^i)^2) D((k_n^i)^2)}
\,
\left[{k}_{n+}^i\cdot {k}_{n}^i \pm M((k_{n+}^i)^2)M((k_n^i)^2))\right].
\label{Pi_pi}
\end{eqnarray}
The mesonic contributions to the thermodynamic potential are then
(to lowest order) given by the ring sum
\cite{NJLMott},
\begin{eqnarray}
\label{Om_corr}
\Omega_{\mathrm{corr}}(T)&=&
\sum \limits_{M=\pi,\sigma} \frac{d_M}{2} \int_{p,m}
\ln \left[ 1 - G \Pi_M(\vec{p},\nu_m)\right]
\end{eqnarray}
with the mesonic degeneracy factor $d_M$.

The fermionic Matsubara sum, the three-momentum integration and
the sum over the three color modes $i=0,+,-$ in (\ref{Pi_pi}) are evaluated
numerically with a finite result for $\Pi_\pi(\vec{p},\nu_m)$ to be inserted
in Eq. (\ref{Om_corr}).
The bosonic Matsubara sum in (\ref{Om_corr}) and the three momentum integral
are also performed numerically.

\begin{figure}[h]
\setcaptionmargin{5mm}
\onelinecaptionsfalse 
\resizebox{.4\textwidth}{!}{\includegraphics{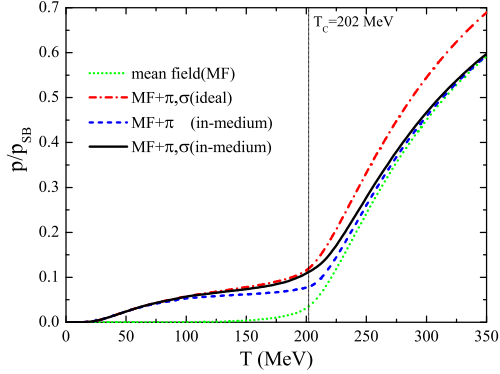}}
\captionstyle{normal} \caption{
Scaled pressure $p/p_{SB}$ in the nonlocal PNJL model with the physical pion 
mass:
mean field contribution (green dotted line),
mean field $+$ pion (blue dashed line),
mean field $+$ pion $+$ sigma (black solid line).
The red dash-dotted line denotes the scaled pressure of an ideal
pion $+$ sigma gas with fixed masses.}
\label{press}
\end{figure}

Our model predictions for the the pressure $P(T) = -\Omega(T)$,
divided by the Stefan-Boltzmann limit, are displayed in Fig.~\ref{press}.
For comparison with the full result
we also show the mean-field result and the
mean-field plus pion contribution as well as the
result for an ideal pion and sigma gas with the masses fixed at
their vacuum values.
We find that
at low temperatures the mean-field (i.e., quark) contribution is suppressed
and the pressure can be well described by a free pion gas.
Near the critical temperature the $\sigma$ meson gives an additional visible
contribution whereas already after $T>1.5~T_c$ the mesonic contributions are
negligible and the quark-gluon mean-field dominates the pressure.

\begin{figure}[h]
\setcaptionmargin{5mm}
\onelinecaptionsfalse 
\resizebox{.4\textwidth}{!}{\includegraphics{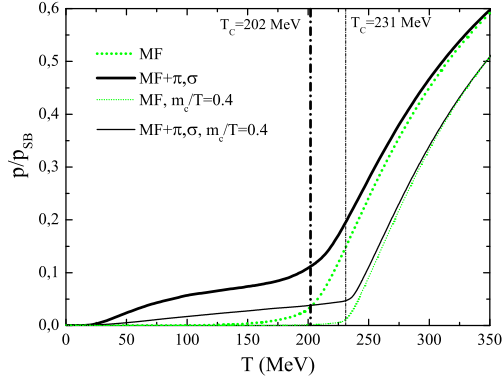}}
\captionstyle{normal} \caption{
Scaled pressure $p/p_{SB}$ in the nonlocal PNJL model with the
physical pion mass (thick lines) and with $m_c/T=0.4$ (thin lines):
mean field contribution (green dotted line),
mean field $+$ pion $+$ sigma (black solid line).}
\label{pressMcDivT04}
\end{figure}


So far we have used the parameter set of  Ref.~\cite{GomezDumm:2006vz}
with $m_c = 5.8$~MeV, corresponding to the physical pion mass of 140~MeV 
in vacuum. 
On the other hand, in most lattice calculations the masses are considerably 
larger. Thus in order to perform a meaningful comparison with lattice
results, we should repeat our model calculations using quark masses
similar to the lattice ones. 
To be specific, we choose the current quark mass to scale with the
temperature as $m_c = 0.4~T$, mimicking the situation in the lattice
calculation of Ref.~\cite{Karsch:2000ps}, where the up and down quark
masses behave in the same way.

Our results for the scaled pressure are displayed in Fig.~\ref{pressMcDivT04}.
For comparison we show again the results obtained with $m_c = 5.8$~MeV 
(thick lines). 
With $m_c = 0.4~T$ (thin lines) the qualitative behavior remains unchanged:
Above $T_c$ the total pressure (black solid line) converges quickly to the 
mean-field result (green dotted line), whereas below $T_c$ the pressure is
dominated by the mesonic contribution.
Quantitatively, the meson contributions are of course strongly suppressed
in the case of the heavier quark masses.
Also note that the mean-field result is affected as well. In particular
$T_c$ rises from 202~MeV for $m_c = 5.8$~MeV to 231~MeV for  $m_c = 0.4~T$

\begin{figure}[h]
\setcaptionmargin{5mm}
\onelinecaptionsfalse 
\resizebox{.4\textwidth}{!}{\includegraphics{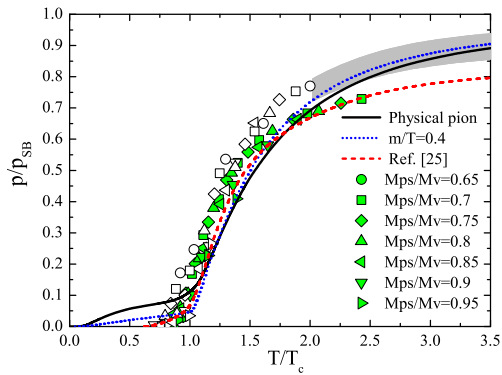}}
\captionstyle{normal} \caption{
Scaled pressure $p/p_{SB}$ as a function of $T/T_c$:
non-local PNJL model with physical pion mass (black solid line)
and with $m_c/T=0.4$ (blue dotted line).
Red dashed line: Lattice data for two-flavor QCD with
staggered quarks ~\cite{Karsch:2000ps}. 
The shaded region is an estimated continuum extrapolation of these data
for massless QCD \cite{Karsch:2000ps}.
Points: Lattice data for two-flavor QCD with
Wilson-type quarks ~\cite{AliKhan:2001ek} for $N_t=6$ (open symbols) and
$N_t=4$ (green filled symbols). 
The data for the pressure \cite{AliKhan:2001ek}
have been divided by the Stefan-Boltzmann limit for $N_t=6$ and $N_t=4$, 
respectively, as given in Ref.~\cite{AliKhan:2001ek}.}
\label{pressLat}
\end{figure}

In Fig.~\ref{pressLat} we compare our model results for the scaled pressure
with lattice simulations.
Our model results are indicated by the black solid line and the blue
dotted line, which correspond to the physical parameterization 
($m_c = 5.8$~MeV) and the case $m_c = 0.4~T$, respectively. Both
curves include mesonic correlations and are identical to the black solid
lines in Fig.~\ref{pressMcDivT04}. Note, however, that they are now
displayed as functions of $T/T_c$, with $T_c$ being different for the
two parameterizations.

The red dashed line indicates the result of Ref.~\cite{Karsch:2000ps}
obtained on a $16^3\times 4$ lattice with improved staggered fermion actions
for two flavors. As mentioned above, in this calculation the quark mass is 
unphysically large and scales with the temperature, $m_{u,d}/T = 0.4$.
At $T \gtrsim 2T_c$ our results are systematically above the lattice
data. Since the latter are obtained with four time slices only ($N_t = 4$),
this is probably a lattice artifact.
This interpretation is supported by the lattice results of
Ref.~\cite{AliKhan:2001ek} for Wilson fermions with $N_t = 4$
(green filled symbols) and $N_t = 6$ (open symbols).
While the former almost coincide with the results of
Ref.~\cite{Karsch:2000ps}, the latter indicate higher pressures at
large $T$. Although the Wilson data have been extracted
on ``lines of constant physics'', i.e.,
for $T$-independent quark masses
and are therefore not entirely comparable to the staggered fermion data,
this tendency should be correct.
Indeed, the authors of Ref.~\cite{Karsch:2000ps}
tried a continuum extrapolation for massless QCD (shaded area), which
is in fair agreement with our model results for both, physical and large
quark masses.

Whereas in our model, even for $m_c = 0.4~T$, there remains a visible pion 
contribution to the pressure down to $T/T_c \approx 0.2$, 
the lattice pressure of \cite{Karsch:2000ps} vanishes at $T/T_c = 0.6$.
This discrepancy has a trivial explanation by the fact that the lattice
pressure has been obtained by the ``integral method'' which leaves one
integration constant undetermined.
In Ref.~\cite{Karsch:2000ps}, this constant has been fixed by the choice
that the pressure vanishes at $T/T_c = 0.6$.
On the other hand, chiral perturbation theory predicts that at very low
temperature the pressure is well described by an ideal pion gas,
in good agreement with our model.
Hence, if the integration constant on the lattice had been fitted to
$\chi PT$, rather than setting the pressure at $T = 0.6~T_c$ equal to
zero, they would be in good agreement with our results at this point.
On the other hand, our approach underestimates the lattice
data in the region $0.9 \lesssim T/T_c \lesssim 1.6$.
This may be attributed to our neglect of
hadronic resonances, other than pion and sigma.

In summary, we have extended previous studies of the thermodynamics of
low-energy QCD within Polyakov-loop NJL models to include mesonic
correlations.
To that end, we have evaluated the pion and sigma contributions to
the pressure by calculating the ring sum. We have thereby employed
a nonlocal four-fermion interaction of the instanton liquid type.
The pionic contribution dominates the pressure in the low temperature
region. In this regime, the pressure is quite insensitive to the details
of the interaction and agrees almost exactly with that of an ideal pion gas.
At temperatures $\gtrsim T_c$, on the other hand, the mesonic contributions
die out.
For unphysical pion masses, as obtained in most present lattice calculations,
mesonic correlations play only a minor role, even below $T_c$,
in accordance with the lattice results.

For the future we plan to include the  back reaction effect
of mesonic correlations to the equation of motion of mean fields.
This can lead to a lowering of $T_c$.
We also plan to study nonzero chemical potentials,
which may require a nontrivial extrapolation of the Polyakov-loop
potential into this regime \cite{Schaefer:2007pw}. We then should
include baryonic degrees of freedom as well.

\begin{acknowledgments}
We thank O.~Kaczmarek, C. Sasaki, and K. Redlich for critial
remarks and illuminating discussions.
A.E.R. thanks J.~Wambach for his hospitality at TU Darmstadt.
We acknowledge support by the Heisenberg-Landau programme
(M.B., A.E.R., M.K.V), by the Russian Foundation for Basic Research
under contract 05-02-16699 (A.E.R., M.K.V), by BMBF (A.E.R)
and by the Polish Ministry of Science and Higher Education (D.B.).
\end{acknowledgments}

\newpage
\selectlanguage{russian}
\begin{center}
\large \bfseries \MakeTextUppercase{%
Эффект мезонных корелляций в фазовом переходе КХД}
\end{center}
\begin{center}
\bfseries Д.~Блашке, М. Бубала, А.Е.~Раджабов, М.К.~Волков
\end{center}
\begin{center}
\begin{minipage}{\textwidth - 2cm}
\small
Фазовый переход сильновзаимодействующей материи при конечной температуре изучен в нелокальной киральной кварковой модели типа НИЛ соединенной с петлей Полякова. В отличие от существующих исследований в рамках среднего поля в модель включены мезонные корреляции с помощью кварк-антикварковой кольцевой суммы. Для физических масс пиона нами обнаружено, что пионы дают доминирующий вклад в давление, тогда как выше $T_c$ давление хорошо описывается результатом вычислений в приближении среднего поля. Для большой массы пиона, как в вычислениях на решетке, мезонные эффекты подавлены.
\end{minipage}
\end{center}
\selectlanguage{english}
\end{document}